\documentclass[%
 reprint,
superscriptaddress,
 amsmath,
 amssymb, 
 aps,  
]{revtex4-2}

\usepackage{graphicx}
\graphicspath{{plots/fig1/}{plots/fig2/}{plots/fig3/}{PDF_output/}{Final_plots/}}
\usepackage{comment}
\usepackage{dcolumn}
\usepackage{bm}
\usepackage{hyperref}

\usepackage{xfrac}
\usepackage{siunitx}
\usepackage[T1]{fontenc} 
\usepackage[dvipsnames]{xcolor}
\usepackage{ulem}
\usepackage{tikz}
\usepackage{upgreek}

\colorlet{ForestGreen}{Black}
\begin{document}

\preprint{APS/123-QED}
\title{Single V2 defect in 4H Silicon Carbide Schottky diode at low temperature}

\author{Timo Steidl}
\thanks{These two authors contributed equally}
 \affiliation{3rd Institute of Physics, IQST, and Research Center SCoPE, University of Stuttgart, Stuttgart, Germany}

\author{Pierre Kuna}
\thanks{These two authors contributed equally}
 \affiliation{3rd Institute of Physics, IQST, and Research Center SCoPE, University of Stuttgart, Stuttgart, Germany}

\author{Erik Hesselmeier-Hüttmann}
 \affiliation{3rd Institute of Physics, IQST, and Research Center SCoPE, University of Stuttgart, Stuttgart, Germany}

 \author{Di Liu}
 \affiliation{3rd Institute of Physics, IQST, and Research Center SCoPE, University of Stuttgart, Stuttgart, Germany}
 
 \author{Rainer St\"ohr}
 \affiliation{3rd Institute of Physics, IQST, and Research Center SCoPE, University of Stuttgart, Stuttgart, Germany}

  
\author{Wolfgang Knolle}
\affiliation{Department of Sensoric Surfaces and Functional Interfaces, Leibniz-Institute of Surface Engineering (IOM), Leipzig, Germany}

\author{Misagh Ghezellou}
\author{Jawad Ul-Hassan}
\affiliation{Department of Physics, Chemistry and Biology, Linköping
 	University, Linköping, Sweden}

\author{Maximilian Schober}
\affiliation{Institute for Theoretical Physics, Johannes Kepler University Linz, Linz, Austria}
\author{Michel Bockstedte}
\affiliation{Institute for Theoretical Physics, Johannes Kepler University Linz, Linz, Austria}

\author{Adam Gali}
\affiliation{HUN-REN Wigner Research Centre, Budapest, Hungary}
\affiliation{Institute of Physics, Department of Atomic Physics, Budapest University of Technology and
Economics, Budapest, Hungary}
\affiliation{MTA-WFK Lendület “Momentum” Semiconductor Nanostructres Research Group}

\author{Vadim Vorobyov}
\email[]{v.vorobyov@pi3.uni-stuttgart.de}
 \affiliation{3rd Institute of Physics, IQST, and Research Center SCoPE, University of Stuttgart, Stuttgart, Germany}  

\author{J\"org Wrachtrup}
 \affiliation{3rd Institute of Physics, IQST, and Research Center SCoPE, University of Stuttgart, Stuttgart, Germany}
 \affiliation{Max Planck Institute for solid state physics, Stuttgart, Germany}


\begin{abstract}
Nanoelectrical and photonic integration of quantum optical components is crucial for scalable solid-state quantum technologies. 
Silicon carbide stands out as a material with mature quantum defects and a wide variety of applications in semiconductor industry.
Here, we study the behaviour of single silicon vacancy (V2) colour centres in a metal-semiconductor (Au/Ti/4H-SiC) epitaxial wafer device, operating in a Schottky diode configuration. 
We explore the depletion of  free carriers in the vicinity of the defect, as well as electrical tuning of the defect optical transition lines. 
By detecting single charge traps, we investigate their impact on V2 optical line width. 
Additionally, we investigate the charge-photon-dynamics of the V2 centre and find its dominating photon-ionisation processes characteristic rate and wavelength dependence. 
Finally, we probe the spin coherence properties of the V2 system in the junction and demonstrate several key protocols for quantum network applications. 
Our work shows the first demonstration of low temperature integration of a Schottky device with optical microstructures for quantum applications and paves the way towards fundamentally scalable and reproducible optical spin defect centres in solids. 
\end{abstract}

\maketitle

\section{Introduction}
Over the recent decades, colour centres have matured to a valuable resource for quantum applications \cite{Awschalom_2018}.
However, hosted in solids they are subject to the fluctuating electron charge environment \cite{orphal2023optically, Pont2021} which deteriorates their quantum properties.
Elimination of charge fluctuations in the host material is key for fundamental scalability of the platform.
Silicon carbide (SiC) is a mature and well developed semiconductor material system \cite{sciencesic}.
Due its unique combination of optical, semiconductor and material properties it is an excellent host for quantum emitters and spin defects \cite{Ou_2024}. 
Importantly, the ability to control the Fermi level, i.e. the charge environment of defects by controlled doping is advantageous for integrating optically addressable solid state defects.
For example, in-plane Schottky junctions in diamond, in conjunction with surface termination were shown to be effective for defect charge state switching \cite{Schreyvogel_2016}.  
In SiC, integration of defect ensembles with PIN junctions was reported \cite{widmann2019electrical}, leading to electrically detected magnetic resonance, and precise charge control.
Recently, p$^{+}$-p-n{$^+$} structures \cite{day2024electrical} were tested for integration with G centres in silicon on an insulator material platform. 
For quantum optical applications the optical emission linewidth is a key parameter. Spectrally narrow and stable emission lines enable e.g large entanglement rates in quantum repeater networks. 
Previously, PIN junctions were studied at temperatures of a few Kelvin in commercially available SiC samples, for precise charge control of di-vacancy (VV$^0$) centres \cite{anderson2019electrical}. 
Usually, the application of a negative bias to deplete charges in the intrinsic region of the PIN structure leads to the creation of defect in an ionised charge states which is not emitting photons \cite{delasCasas2017}.
While the  silicon vacancy (V2) defect in 4H-SiC is a promising scalable spin photon interface platform \cite{hesselmeier2024high}, \cite{babin2022fabrication} with coherent optical transitions up to $T= \SI{20}{K}$ \cite{Udvarhelyi2020}.
no single V2 centers have not been investigated in semiconductor junctions. \cite{widmann2019electrical, Bathen2019}.


\begin{figure*}
	\centering
	\includegraphics[width=\textwidth]{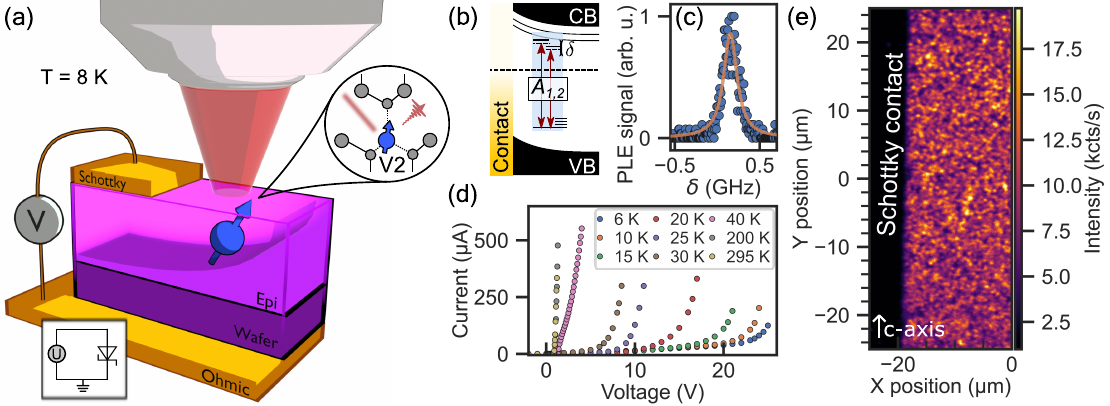}
	\caption{V2 colour centre in Schottky diode: 
	\textbf{(a)} Schematic view of V2 colour centre with depletion zone of charge carriers formed next to the barrier.
	\textbf{(b)} Schematic band gap bending diagram in the vicinity of the junction with mid gap defect states. Allowed transitions $A_{1,2}$ with detuning $\delta$. 
	\textbf{(c)} Photoluminescence excitation spectrum (PLE) of single V2 centre with $A_1$ and $A_2$ simultaneously scanned.
	\textbf{(d)} Current-voltage (I-V) characteristic of the obtained Schottky diode, showing the forward biased current, measured at several temperatures from room temperature down to cryogenic environment.
	\textbf{(e)} Confocal map of the studied sample A in the vicinity of the Schottky contact.}
	\label{fig1}
\end{figure*}
In this work we investigate the operation of single silicon vacancy V2 centres in a Schottky diode integrated with optical micro solid immersion lens at low temperature.
We observe that charge depletion yields narrowing of the optical transition lines and furthermore investigate the electrical tuning of the defect via the DC Stark shift. 
By localising the defect optically we precisely map the depletion zone around the Schottky contact and calibrate its local intrinsic doping concentration. 
We find that the defect stays in its negative charge state throughout the range of voltages used in the investigation. 
We support our findings by a theoretical and experimental study of ionisation photo-dynamics of single defects at various excitation wavelengths.
Finally, we show the integration of the Schottky junction with an optical microstructure and demonstrate its use in a typical experimental protocols relevant for the operation of a quantum network, including the measurement-based optical transition stabilisation, electron spin coherent manipulation and nuclear spin repetitive single shot readout. 
With this we combine optical, electrical, electron and nuclear spin control of isolated V2 defects in silicon carbide. 

\begin{figure*}
	\includegraphics[width=\textwidth]{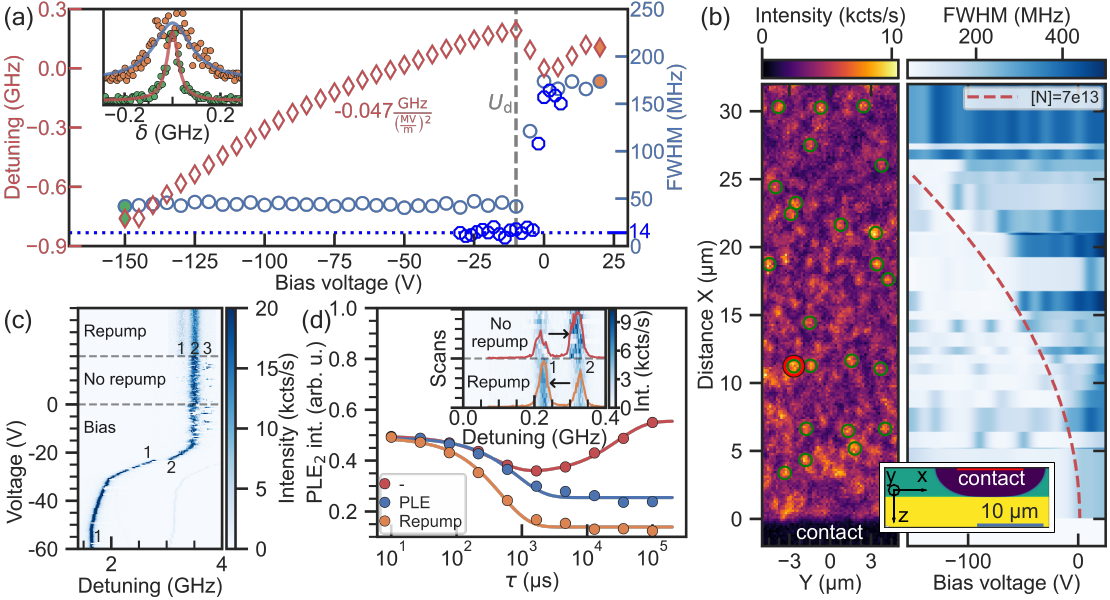}
		\caption{Optical spectroscopy of V2 centre in a Schottky diode:
			\textbf{(a)} Extracted linewidth and spectral PLE position of defect-1 marked with red circle in (b) versus applied voltage (\SI{-150}{V} to \SI{+20}{V}). 
			Detuning follows a purely quadratic dependence yielding perpendicular Stark tuning coefficient of 0.047$\pm$0.006 $\mathrm{GHz}/\left(\frac{\mathrm{MV}}{\mathrm{m}}\right)^2$. 
			Inset show both examples, broad PLE without bias and narrow PLE lines in depleted case (voltage data point indicated with filled markers).
			Blue octagons show measurement of defect-2, which has a lifetime limited (calculation uses lifetimes from Ref. \cite{liu2023silicon}, see SM)
			\textbf{(b)} Right: Boundary of the depletion zone mapped by single defects (shown on the left) and simulated with COMSOL \cite{COMSOL} (red dashed line is the threshold for [e$^{-}$] = \SI{1e12}{cm}$^{-3}$). 
			The COMSOL simulation has been parametrically optimised, best agreement for 2\,µm deep colour centres and initial epi-layer doping concentration (green colour of inset) of [N] = \SI{7e13}{cm}$^{-3}$, validating given sample properties. Substrate doping (yellow colour) is [N] = \SI{1e17}{cm}$^{-3}$.
			\textbf{(c)} Exemplary PLE measurements of defect from sample B with distinct splitting of the resonance intro three equally spaced lines (1,2,3), caused by a nearby charge trap, showing occasional charge trap ionisation and jumps between lines 2-3. Application of off-resonant (repump) laser results in motional averaging and steady state charge distribution between lines 1-2, while the negative bias applied to diode stabilises the charge state of the trap and eventually narrowing down the PLE line. 
			\textbf{(d)} Time transient of single trap charge states next to defect 3 upon no excitation (green),  off-resonance 20\,µW (blue) and resonant to V2 excitation (red) (\SI{3}{nW}). 
			Solid lines are exponential fits.
			}
		\label{fig2}
\end{figure*}

\section{Results}
The V2 centre is a silicon vacancy ($\mathrm{V_{Si}}$) at the cubic lattice site of 4H-SiC epitaxially grown layer. 
It is a very good single photon emitter \cite{nagy2019high} with long spin memory lifetimes \cite{babin2022fabrication}.
In our study we investigate V2 next to a Schottky contact realised by a Au-Ti-SiC interface. 
The electric circuit is closed by semi-insulating silicon carbide (schematically shown in Fig. \ref{fig1}a). 
At cryogenic temperatures, the defect states can be excited efficiently by resonant optical excitation at \SI{1.352}{eV} ($\SI{\sim 916}{nm}$) with spin resolved $A_{1,2}$ optical transitions for $m_{\mathrm{s}}=\pm 1/2, \,\pm3/2$ spin manifolds (Fig. \ref{fig1}b) splitted by \SI{\sim{1}}{GHz} \cite{Udvarhelyi2020}.
We scan both transitions simultaneously with two resonant lasers separated by excited state zero field splitting yielding a single resonance peak in the photoluminescence excitation (PLE) spectrum shown in Fig. \ref{fig1}c.
Exploring the current-voltage (I-V) characteristic of the diode (Fig. \ref{fig1}d).
At low temperatures the diode threshold forward voltage increases compared to room temperature, matching well with the diode law. 
If too much voltage is applied and the current exceeds > mA, heating and electroluminescence can be observed (compare SI Fig. S5).
Using confocal microscopy (Fig. \ref{fig1}e) we localise single defects close to the metal contact.
We perform subsequent PLE measurements at different bias voltages. 
The extracted PLE position and linewidth is shown in Fig. \ref{fig2}a for two defects with defect-1 located $\sim$11.2\,µm from the contact (red circle in Fig. \ref{fig2}b) and defect-2 from another area.
We apply a total voltage sweep from \SI{-150}{V} to \SI{+30}{V}. 
For negative bias, the detuning follows a purely quadratic dependence, with a Stark coefficient which is extracted to be 0.047$\pm$0.006 $\mathrm{GHz}/\left(\frac{\mathrm{MV}}{\mathrm{m}}\right)^2$.
The local effective electric field strength has been determined using a COMSOL \cite{COMSOL} simulation.
As the electric field components point perpendicular to the crystal c-axis this is the first time a perpendicular Stark tuning coefficient for V2 centres in 4H-SiC is reported, with a linear parallel Stark coefficient of 3.7  $\mathrm{GHz}/\left(\frac{\mathrm{MV}}{\mathrm{m}}\right)$ reported in \cite{Lukin2020spectrally}. 
At near-zero or forward bias a PLE spectral broadening is observed. 
Whereas at negative bias, we typically observe a line narrowing from \SI{170}{MHz} down to \SI{40}{MHz}.
The linewidth of defect-1 is measured to be \SI{40}{MHz}, slightly higher then the Fourier limit ($\sim$\SI{20}{MHz}).
The effective lifetime limit of the single PLE curve  will be in between the values of each single optical transition (blue line in Fig. \ref{fig2}a, for calculation see SM).
The most narrow optical line (dark-blue dots in Fig. \ref{fig2}a) is observed at defect-2 and reaches the $A_2$ Fourier limit of around \SI{14}{MHz}. 

The line narrowing effect is reproducible with almost no hysteresis (see SM Fig. S2a). 
PLE narrowing can be explained by the depletion of the surrounding volume charge in the vicinity of the defect. 
The depletion zone size around the Schottky contact rises with larger negative bias voltage (see SM Fig. S2e). 
For the specific defect investigated, the zone in which line narrowing occurs reaches its position at around $U_d=\SI{-10}{V}$, which we denote as depletion voltage. 
However, for the other 30 investigated defects across the sample the depletion voltage varies. 
We combine all data in Fig \ref{fig2}b, where we depict the PLE spectral linewidth depending on the applied voltage and distance from the contact.
The experimentally measured linewidth shows a threshold like behaviour, at voltages predicted by simulations (green dashed line, see SM Fig. S2c-d), once the edge of the depletion zone reaches the defect location, lines get narrow for voltages below depletion voltage. 
From comparing simulation and our experimental results we confirm the estimated doping concentration of the intrinsic layer to be [N$_{\mathrm{Epi}}$] = \SI{7e13}{cm}$^{-3}$. 
The mechanism for line narrowing is best exemplified by single defects with PLE spectral lines further split by strongly coupled nearby electron charge traps (see Fig. \ref{fig2}c).
\begin{figure*}[t]
	\centering
	\includegraphics[width=\textwidth]{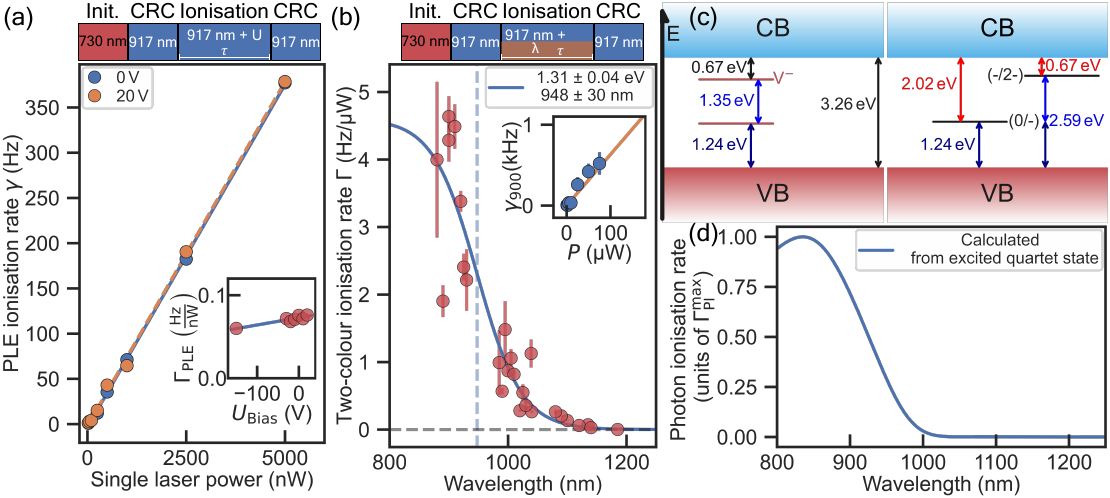}
	\caption{Ionisation behaviour of single V2 centre under PLE and two-colour laser irradiation:
	\textbf{(a)} Ionisation rate $\gamma$ under resonant laser irradiation versus PLE laser power shows almost identical result for two different bias voltages (inset shows line slope $\Gamma$ for various voltages).	
	\textbf{(b)} Two colour absorption power dependent ionisation rate. Second laser pulse (varied in length, wavelength and power) is applied togehter with PLE lasers at low power. 
	Fit with Fermi distribution yields threshold of \SI{948}{nm} for the charge conversion of V$_{\mathrm{Si}}^{-}$ to V$_{\mathrm{Si}}^{2-}$, compare (c).
	\textbf{(c)} Defect level and charge state transition level picture of the cubic silicon vacancy center within the bandgap (\SI{3.26}{eV} used).
	Values for the (-/2-)-transition taken from Ref. \cite{widmann2019electrical}, the other values are deduced from that. 
	\textbf{(d)} \textit{Ab-initio} calculation of the photon absorption cross section for the charge transition ($-/2-$) of V2, recalculated to a photon ionisation rate $\Gamma_{\mathrm{PI}}$ (more detailed information in SM).
	Similar wavelength dependence visible like measured data in (b). 
	}
	\label{fig3}
\end{figure*}
The figure shows a PLE spectrum of a single defect with three PLE lines. 
We confirm that each line belongs to the same defect, by performing hyperfine resolved optical detected magnetic resonance (ODMR), see Fig. S4h. 
We attribute this behaviour to the presence of a single nearby fast switching charge trap, producing a discrete Stark shift of the defect \cite{ji2024correlated, delord2024correlated}. 
As we see three PLE positions (Fig. \ref{fig2}c lines 1-3) we assume the charge trap to be a carbon vacancy $\mathrm{V_C}$ as it has three stable charge states $(2-,-,0)$ \cite{Son2021}. 
The charge trap dynamics can be accelerated by laser illumination \cite{Wolfowicz2017} leading to motional averaging and redistribution of its steady state population distribution \cite{Pont2021}. 
For zero bias voltage and no additional laser we observe that the V2 PLE is mostly found in the centre position (line 2), showing only rare shifts to other spectral positions (line 3). 
Applying a weak repumping laser during the PLE scan partly populates the charge trap into the third charge state (line 1).
The population in line 3 can not be observed under this condition. 
As soon as the depletion zone reaches the defect (causing a jump in the PLE and a linewidth narrowing) the charge trap is depleted so that just one line remains visible. 
For another V2 defect, two PLE positions can be observed at zero bias (inset of Fig. \ref{fig2}d). 
In the following, we explore the temporal dynamics of the trap charge states. 
We applied the following sequence:  a charge resonance check (CRC1) at the right line, then repumping, either with a resonant laser or just wait for a certain time and finally another resonance check at the right line (CRC2). 
One can observe a depopulation if using a laser pulse and a recovery if no laser is applied. 
The timescale of charge transfer is on the order of a few milliseconds.
While the traps switch the charge state within the depletion zone the V2 defect doesn't change its charge state dynamics at the applied voltage bias (see Fig. S3a). 
This is different for other defects integrated into semiconductor devices \cite{delasCasas2017, day2024electrical}. 

\begin{figure*}[t]
	\centering
	\includegraphics[width=\textwidth]{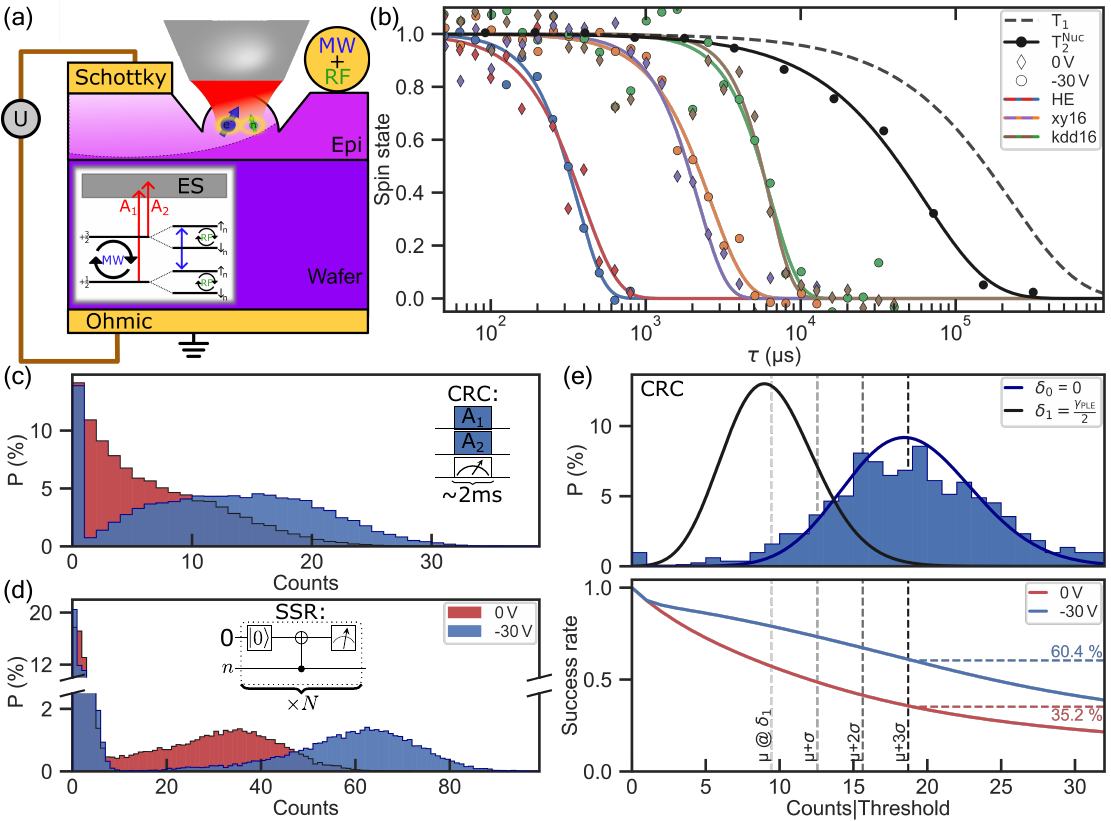}
	\caption{Combination of electrical and optical structures and observation of spin properties under negative bias:
	\textbf{(a)} Schematic drawing of a sample showing a colour centre integrated in a SIL within the depletion zone of the Schottky contact. Inset: Ground state level structure with spin hyperfine splitting.
	\textbf{(b)} Electron and nuclear spin memory coherence measured with various dynamical decoupling sequences. Dashed line show $T_1$ limit, fitted by stretched exponential decays.
	\textbf{(c)} Charge resonant check histogram for \SI{0}{V} and \SI{-30}{V} accumulated over several hours, showing higher count rates for depleted case. 
	\textbf{(d)} Single shot readout histogram for both voltages. Using the more distinct Poissonian distribution for negative bias allows higher measurement rates by more favourable pre-selecting thresholds.
	\textbf{(e)} Top: blue histogram shows best CRC measurement which can be attributed to the best frequency overlap of resonant laser and optical transition.
	Simulation of CRC measurement with a laser detuned by the defects natural halfwidth half-maximum (HWHM).
	Bottom: success rate is the fraction of the total SSR events which will be used as a function of set collection threshold for normal and depleted case.
	Dashed lines show first and second moment of photon counts distribution of defect detuned by half width half maximum.
	Depleted colour centres yield a measurement fidelity of more than \SI{99}{\%} ($\mu + 3 \sigma$) without losing more than the half of SSR events.
	}
	\label{fig4}
\end{figure*}

To further gain insight, we perform ionisation rate measurements under various wavelengths and laser powers. 
Fig. \ref{fig3}a shows the power dependent ionisation rate under resonant excitation only. 
Measurements have been taken at several bias voltages showing no significant change, indicating that the underlying processes are independent of the defects charge environment. 
Fig. \ref{fig3}b shows the ionisation rate for a case where we apply an additional laser pulse together with the resonant laser. 
Using two adjustable single mode lasers (Toptica CTL,  Hubner C-Wave) we were able to use different wavelengths between \SI{880}{nm} and \SI{1200}{nm} for redistribution of charges in the vicinity of the defect (see Fig.3). 
Each point marks the slope of a power dependent ionisation rate. 
There is a significant drop for wavelengths beyond \SI{950}{nm} which corresponds to a threshold energy for ionisation of \SI{1.31}{eV}. 
This perfectly matches the \textit{ab-initio} calculations for the ionisation cross section from the negative charge to the double negatively charged state $(-) \to (2-)$ shown in Fig. \ref{fig3}d.
Discrepancies between the measured ionisation rate and calculated values ($\Gamma_{\mathrm{PI}}$) might occur due to power calibration (measured outside the cryo versus expected intensity at the defect) or no perfect occupation of the excited state and competing processes like repumping.
While ionisation to the neutral state is also possible, theory predicts the tenfold smaller rate compared to our experimental observation.
Additionally, the defect would transform from the V$^0$ state immediately back to the bright charge state under  resonant excitation at \SI{1.352}{eV}. 
From this we can infer that the defect predominantly ionises into the dark state V$^{2-}$. 
Depletion of the electron density in the vicitinity of the defect thus further reduces the probability and prevents the defect from beeing in the dark state, evident from reduced ionisation rate (inset of Fig. \ref{fig3}a) and confocal maps at negative bias (see SM).  

Next we demonstrate the integration of the Schottky diode and a microscopic solid immersion lens (SIL).
The same sample has recently been used to show nuclear assisted single shot readout \cite{HesselmeierQudit, hesselmeier2024high}. 
The main motivation here was to confirm that electrical and optical structures can be combined (Fig. \ref{fig4}a) leading to a tuneable spin photon interface platform based on V2 centres in SiC.
The challenge here is to ensure that the defect located in the depletion zone, while not deteriorating the quality of the optical structure by metallic contacts, and ensuring the excellent spin coherence properties of the defect (see SM).

Similar to previously reported, the defect electron (nuclear) spin, show \SI{0.4}{ms}(\SI{70}{ms}) of Hahn echo $T_2$ time, which can be further extended via the dynamical decoupling \cite{de_Lange_2010} (see SM Fig. S4) limited by magnetic noise from a bath with natural abundance of nuclear spin isotopes. 
By applying the negative bias, we confirm that those parameters are preserved within the depletion zone of the junction, crucial for applications. 

\section{Discussion}
The effect of the optical line narrowing has its most implications on optical readout and emitted coherent photons quality. 
Its importance is evident from the clear improvement in the resonant charge state readout (CRC) and nuclear spin repetitive readout (SSR) photon statistics presented in Fig. \ref{fig4}c-d. 
For the charge state readout, the raw photon statistics accumulated over several hours of measurement time shows a well defined Poissonian lobe for the case of $\mathrm{U}=\SI{-30}{V}$ of bias voltage, while showing a decaying distribution for unbiased case, caused by significant spectral diffusion. 
For the readout of the nuclear spin state, due to the better stabilised optical line position, during the lengthy measurements, the figure demonstrates a brighter and better defined photon histograms with a drastically change in the middle overlapping region for different quantum states Fig. \ref{fig4}d.

It is crucial for spin measurements and spin photon entanglement protocols, that an optical transitions would not spectrally shift (\textit{diffuse}) during the whole measurement run. 
Alternatively, the excitation and emission rates of defects would drift causing an infidelity in calibrated parameters. 
Furthermore, with spin photon entangled state, the spectral diffusion leads to unwanted phase accumulation on the photons, leading the dephasing of the entangled state. 
Normally, for solid state defects, it is ensured by applying a charge-resonance check measurement, which probes the fluorescence level.
When number of photons exceeds a certain threshold, defect resonance is considered to be within a small interval centred around the laser frequency with high degree of confidence.  
As a typical example, here we find the success rate of finding a defect resonance within its natural halfwidth interval around the laser frequency and find that with confidence of 99.85\% (above $\mu+3\sigma$) at the depleted case 60.4 \% of the entanglement attempts would be successful while for the default case it is only 35\%. Fig. \ref{fig4}e bottom panel. 
This could be even more important for samples with larger intrinsic nitrogen concentration and is affecting the entanglement link efficiency directly, as less entanglement attempt per channel is require for a success. 

In conclusion, our findings mark novel insights into the charge state dynamics of the V2 centre, underlining the application relevant negatively charged state stability \cite{Ou_2024}. 
These observations make combination of V2 cantre with electronic structures in SiC a promising system for quantum network applications and spin photon interface \cite{kalb2017entanglement}. 
The observation of discrete charge traps opens a new possibility for the optical addressability of dark colour centres, potentially suitable for forming a larger spin cluster around the V2 system \cite{ji2024correlated, delord2024correlated}. 
Our results pave the way for the integration of junctions with nano beams which promise high collection efficiencies \cite{Krumrein2024Precise, Tiecke2015Efficient} (see SM) and unleashes the full potential of the material platform.

\acknowledgments
P.K., J.U.H., and J.W. acknowledge support from the European Commission through the QuantERA project InQuRe (Grant agreements No. 731473, and 101017733).
P.K. and J.W. acknowledge the German ministry of education and research for the project InQuRe (BMBF, Grant agreement No. 16KIS1639K).
J.W. acknowledge support from the European Commission for the Quantum Technology Flagship project QIA (Grant agreements No. 101080128, and 101102140), the German ministry of education and research for the project QR.X (BMBF, Grant agreement No. 16KISQ013) and Baden-Württemberg Stiftung for the project SPOC (Grant agreement No. QT-6).
J.W. also acknowledges support from the project Spinning (BMBF, Grant agreement No. 13N16219) and the German Research Foundation (DFG, Grant agreement No. GRK2642).
J.U.H. further acknowledges support from the Swedish Research Council under VR Grant No. 2020-05444 and Knut and Alice Wallenberg Foundation (Grant No. KAW 2018.0071).
M.S. and M.B. acknowledge financial support from the Austrian Science Fund (FWF, grant I5195) and German Research Foundation (DFG, QuCoLiMa, SFB/TRR 306, Project No. 429529648).

\appendix
\section{Experimental setup}
All experiments were performed at cryogenic temperature $<$\SI{10}{\kelvin} (if not stated differently) in a Montana Instruments cryostation. 
A self-build confocal microscope was used to optically excite single V2 centres and detect the red-shifted phonon side band. 
Initialisation of the charge state is performed by off-resonant excitation via a \SI{728}{nm} diode laser (Toptica iBeam Smart).
For resonant optical excitation we used an external cavity tuneable diode laser (Toptica DL Pro), which was split and frequency shifted by two separate AOMs to address both optical transitions selectively. 
Those transitions are called A1 and A2 and are split by $\approx$\SI{1}{\giga\hertz}, depending on the zero field splitting of the excited state. 
Laser photons are filtered by two tunable long-pass filters (Semrock TLP01-995).
The magnetic field is created via an electromagnet from GWM Associates (Model 5403EG-50) connected to a power supply from Danfysik (SYSTEM 8500 Magnet Power Supply). In this work a magnetic field of $B\sim0.21$\,T was used for measurements at sample B.
If not indicated otherwise, we used \SI{20}{nW} A2 excitation power before the cryostation.
The used detectors are fibre coupled superconducting nanowire single-photon detectors from Photon Spot.

\section{Silicon carbide samples}
Sample A had 10\,µm epitaxial layer on 4H-SiC substrate and has natural abundance of silicon (4.7 \% $^{29}$Si) and carbon (1.1 \% $^{13}$C) isotopes, which are spin $I = 1/2$ nuclei. 
Samples were electron irradiated (\SI{5}{kGy}, \SI{5}{MeV}) and annealed at \SI{600}{\degree C} in \SI{850}{mbar} Argon atmosphere.
1\,µm width waveguides have been created by angled RIE.
Sample B is almost identical besides of an implantation dose of \SI{20}{kGy}. 
For the fabrication of the solid immersion lenses a gallium FIB machine was used.
Schottky contacts have been created by masked electron-beam physical vapour deposition of gold with a thin adhesion layer in between.
The bottom ohmic contact is formed by the contact of silver paste with the highly doped SiC substrate. 
 
%
 


\end{document}


\preprint{APS/123-QED}

\title{Supplementary Information: Single V2 defect in 4H Silicon Carbide Schottky diode at low temperature}

\author{Timo Steidl}
\thanks{These two authors contributed equally}
 \affiliation{3rd Institute of Physics, IQST, and Research Center SCoPE, University of Stuttgart, Stuttgart, Germany}

 \author{Pierre Kuna}
 \thanks{These two authors contributed equally}
 \affiliation{3rd Institute of Physics, IQST, and Research Center SCoPE, University of Stuttgart, Stuttgart, Germany}
 
 \author{Erik Hesselmeier-Hüttmann}
 \affiliation{3rd Institute of Physics, IQST, and Research Center SCoPE, University of Stuttgart, Stuttgart, Germany}
 
 \author{Di Liu}
  \affiliation{3rd Institute of Physics, IQST, and Research Center SCoPE, University of Stuttgart, Stuttgart, Germany}

\author{Rainer St\"ohr}
\affiliation{3rd Institute of Physics, IQST, and Research Center SCoPE, University of Stuttgart, Stuttgart, Germany}

  
\author{Wolfgang Knolle}
\affiliation{Department of Sensoric Surfaces and Functional Interfaces, Leibniz-Institute of Surface Engineering (IOM), Leipzig, Germany}

\author{Misagh Ghezellou}
\author{Jawad Ul-Hassan}
\affiliation{Department of Physics, Chemistry and Biology, Linköping
 	University, Linköping, Sweden}
\author{Maximilian Schober}
\affiliation{Institute for Theoretical Physics, Johannes Kepler University Linz, Linz, Austria}
\author{Michel Bockstedte}
\affiliation{Institute for Theoretical Physics, Johannes Kepler University Linz, Linz, Austria}
\author{Adam Gali}
\affiliation{HUN-REN Wigner Research Centre, Budapest, Hungary}
\affiliation{Institute of Physics, Department of Atomic Physics, Budapest University of Technology and
Economics, Budapest, Hungary}
\affiliation{MTA-WFK Lendület “Momentum” Semiconductor Nanostructres Research Group}

\author{Vadim Vorobyov}
\email[]{v.vorobyov@pi3.uni-stuttgart.de}
 \affiliation{3rd Institute of Physics, IQST, and Research Center SCoPE, University of Stuttgart, Stuttgart, Germany}  

\author{J\"org Wrachtrup}
 \affiliation{3rd Institute of Physics, IQST, and Research Center SCoPE, University of Stuttgart, Stuttgart, Germany}
 \affiliation{Max Planck Institute for solid state physics, Stuttgart, Germany}


\maketitle

\section{Measurements of PLE linewidth}
\begin{figure*}[t]
	\centering
	\includegraphics[width=\textwidth]{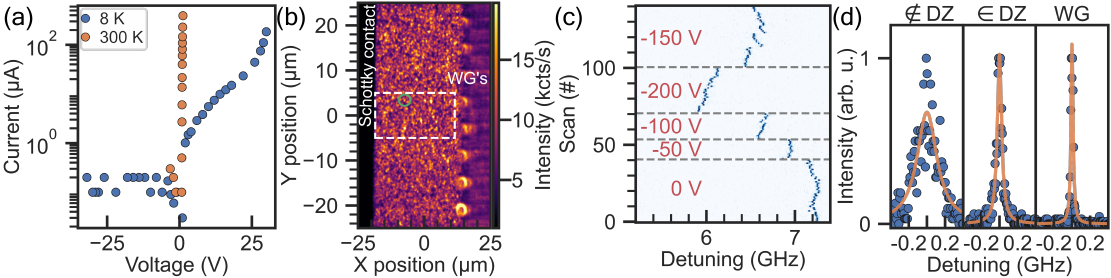}
	\caption{
		\textbf{(a)} Diode characteristics for two different temperatures between room and cryogenic temperatures, also showing the small reverse saturation current.
		\textbf{(b)} Full confocal map showing  the position of the defects from the statistics in Fig 2 in the main text (dashed white rectangular) and location of the waveguides. The defect marked with the green circle was used for the repetitive PLE measurement in Fig \ref{figS2}a.
		\textbf{(c)} Subsequent PLE measurements of the V2 centre in a waveguide. 
		Each block has fixed applied bias voltage and the jumps at the boundaries are caused by Stark shift as we change the voltage and thus also the electric field.
		Within each block the voltage is kept constant, so every jump or drift is related to surface charge fluctuations. 
		Higher negative voltage seems not to improve defect stability.
		\textbf{(d)} Comparsion of PLE measurement of defects within ($\gamma_{\in }$ = 56.7$\pm$\SI{1.3}{MHz}) and without ($\gamma_{\notin }$ = 267.4$\pm$\SI{14.3}{MHz}) the depletion zone as well as integrated into a nanophotonic beam. 
		PLE narrowing by depletion is quite clear. 
		Suprisingly, defects in waveguides appear even more narrow ($\gamma_{\mathrm{WG}}$ = 27.2$\pm$\SI{0.5}{MHz}).
		}
		\label{figS1}
\end{figure*}

The voltage we can apply for Stark tuning or linewidth narrowing is limited by the I-V-characteristics (Fig \ref{figS1}a \& Fig. 1d in the main text) of the diode.
In reversed bias we have nearly no limit as no or just very little reversed current is created.
For positive voltages, the threshold voltage luckily rises for lower temperature, that gives a larger voltage range for forward voltage bias at cryogenic temperatures to operate without sample heating and increasing the linewidths caused by current flow.
Too large current will also cause electroluminescence (see Fig. \ref{figS5}) or start degradation of the contacts.
\newline
As we are using the two-laser PLE scheme for all measurements, we can not extract both single linewidths for $A_1$ and $A_2$, instead we see a superposition of both linewidths.
Both processes sum up and as the $A_2$ transition is typically more dominant, we would expect a PLE linewidth in the same range if we approach the Fourier limit.
For the calculations of the lifetime limits used in the main figure Fig. 2b, where are using the equation $\Delta\nu = 1/(2\pi\tau_{\mathrm{r}})$.
For the excited state lifetimes of $A_{1,2}$ we get $\tau_{A_{1,2}} = 1/(\gamma_{\mathrm{r}}+\gamma_{1,2}+\gamma_{1,2}') = 6.09 | \SI{11.35}{ns}$ using the values from Ref. \cite{liu2023silicon}.
From that we derive $\Delta\nu_{A_{1,2}}$ = 26.1|\SI{14.0}{MHz} as Fourier limits.
To investigate further the natural linewidth of the defect in the depleted area we measure ple linewidth power dependence. 
We measured the PLE linewidth versus resonant laser power for two voltages. 
For the depleted case we typically get an extrapolated linewidth of 40 MHz.
This is already quite close and might be related to some remaining charges as we also found a defect approaching the lifetime limit (main text) 
If we apply positive voltage we get additional broadening from the inserted charges, which offsets the curve in Fig. \ref{figS2}b.
An additional step-like behaviour is visible, which might indicate that even a small amount of resonant light leads to increased charge mobility and subsequent broadening of the PLE lines around the defect. \newline
Not only bulk defects has been studied this work, also defects integrated into 1\,µm wide nano beams.
Fig. \ref{figS1}b show the full confocal scan of Fig. 1e in the main text, here, some of the one-sided tapered waveguides can be seen. 
So, we show PLE measured in a waveguide and observed line shifts from the Stark effect as well.
Each section in Fig. \ref{figS1}c, between the grey dashed lines, are subsequent PLE scans having the same bias voltage applied, which is labeled on the left. 
Smooth voltage scans like in the main text could not be done as the defect would drift faster than the Stark shift would do. 
Using such big voltage steps make the Stark effect quite visible but also drifting and jumping of the PLE lines is still quite recognisable.
The absolute lineshift is also larger than for bulk defects with similar or even shorter distance to the stripline. 
This can be explained as the electric field is screened less, as it is propagating through fewer charge containing material.
As the Schottky contact is to far (> 30\,µm) from the waveguides we did not observe any linewidth narrowing or stabilisation. 
Apparently, we might no be able so see narrowing in nano beams at all as color centers appear already less broad with linewidths of <\SI{30}{MHz} approaching the Fourier limit ($\sim 20$ MHz) (compare Fig. \ref{figS1}d). 
We attribute this to the lack of noticeable volume charge which can lead to broadening. 
On the other hand the defect is closer to the surface of the beam, thus making it more sensitive to surface charges despite the fact that it is still more stable than e.g. NV centers close to the surface. 
This could be an indicator for that SiC is a more comfortable host material as it might has less surface charges in total. 
\newline
In Fig. \ref{figS2}a one can see the alternating PLE position and switching to broad linewidth for near zero voltages. 
In total we swept the voltage 3 times each back and forth between +20 V and -150 Volts. 
Also the repeatability with almost no hysteresis is remarkable but not given for every defect or high repump laser power. 
The data presented in the Fig 2b of main text is extracted from that entiry PLE scan. 
The values from single line fits were averaged over multiple voltage sweeps. 
One might notice some missing data points. 
In these lines the defect was ionised and not successfully repumped. 
It is also notable to say, that the PLE position seems to be tuned in the opposite direction as soon as it is outside of the depletion zone. 
At the edge of the depletion zone some of the charges outside want to go inside again to form equilibrium state and this will cause a build-in field counteracting the one from the contact. 
The defect will then probably be located in between and react according to the sign switching of the electric field.

\section{Numerical simulation of the depletion zone}
\begin{figure*}
	\centering 
	\includegraphics[width=\textwidth]{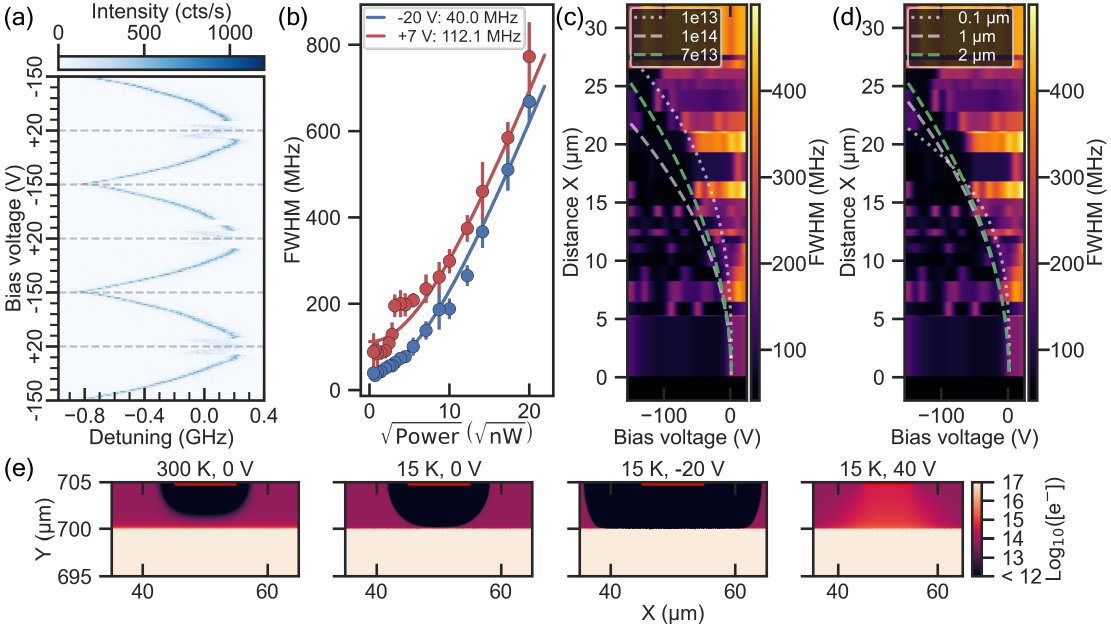}
		\caption{
			\textbf{(a)} Repeated voltage sweeps during PLE used for Fig 2(a) and show reprocudability.
			The detuning of all PLE's in sample A are relative to a resonant laser frequency of 327.123 THz.
			\textbf{(b)} PLE linewidth broadening.
			\textbf{(c)} Depletion zone mapping for different initial epi dopings, 1e13, 1e13
			\textbf{(d)} Depletion zone mapping for different defect depths, 100nm, 1000nm
			\textbf{(e)} COMSOL \cite{COMSOL} simulation of the carrier density for different temperatures and voltages.
			}
		\label{figS2}
\end{figure*}

As in the main text mentioned the simulation of the edge of the depletion zone has been optimized to match with the data of the PLE statistics. 
In the COMSOL simulation we swept the initial intrinsic layer doping concentration as well as the studied defect location depths.
In Fig. \ref{figS2}c \& d it is shown how the depletion zone would occur using the other parameters. 
Partly the difference is not that significant but some values can be excluded as it would cause a different shape.
It is worth to notice again that for each line of a certain distance multiple defect have been used and averaged, depending on the amount of emitters inside a confocal spot and PLE scan range.
The bars of each line have been expanded or narrowed upon the next laying line to close potentially gaps in the data set.
Fig. \ref{figS2}e nicely shows the evolution of the depletion zone for different temperatures and voltages. 
Even at room temperature a large area inside the epi-layer is depleted under the metal contact.
By going down to cryogenic temperatures this area grows and get well defined edges.
The lower the temperature the smaller are the fluctuations, so the simulation done for 15 K also holds for the 8 K, the temperature which has been used for most of our measurements.
Increasing the negative bias shows the same effect. 
Even if it is not visible in this plot, there are also some charges removed from the wafer side, if the temperature is low or the reversed voltage high enough.
But this effect is so small that it does not make an interesting change, the counteracting electric field is also higher there, stopping electrons further to be removed.
It is also shown, that for positive voltages this depletion zone does not only vanish, we also insert additional charges into the epi-layer, which enables current flow and thus could lead to heating and spectral broadening of the optical lines. 
Some details of the setup for the COMSOL simulations are shown in Tab. \ref{tab1}.

\begin{table}[t]
	\centering
	\caption{Used paramter values of the simulation of depletion zone and electric field:}
	\begin{tabular}{|  c |  c |  c |}
		Parameter ($@$\SI{15}{K}) & Value & Unit \\
		\hline\hline
		Workfunction (Au) & 5.1 & eV\\
		\hline
		Workfunction (Ti) & 4.5 & eV\\
		\hline
		Band gap (4H-SiC) & 3.23 & eV\\
		\hline
		Refractive index & 2.588 & 1\\
		\hline
		Electron affinity  & 3.24 & eV\\
		\hline
		Doping concentration (epi) & 7e13 & cm$^{-3}$\\
		\hline
		Doping concentration (wafer) & 1e17 & cm$^{-3}$\\
		\hline
		Eff. density of states, VB & 2.7885e23 & m$^{-3}$\\
		\hline
		Eff. density of states, CB & 1.8881e23 & m$^{-3}$\\
		\hline
		Electron mobility (epi) & 237.02 & m$^2$V$^{-1}$s$^{-1}$\\
		\hline
		Electron mobility (wafer) & 29.556 & m$^2$V$^{-1}$s$^{-1}$\\
		\hline
		Effective mass (e) & 3.6578e-31 & kg\\
		\hline
		Effective mass (h) & 2.4083e-30 & kg\\
		\hline

	\end{tabular}
	\label{tab1}
\end{table}

\section{Charge stability and electroluminescence}
\begin{figure}[t]
	\centering
	\includegraphics[width=\columnwidth]{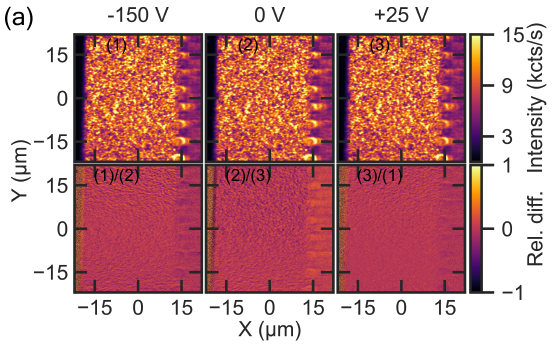}
	\caption{
		\textbf{(a)} Comparsion of confocal scans at three different bias voltages (\SI{-150}{V}/\SI{0}{V}/\SI{+25}{V}). 
		Lower row show subtracted measurements. 
		No significant change can be identified, so so defects appear or disappear by changing the voltage, leading to the assumption, that the defects are charge stable within the voltage range and for off-resonant excitation.
	}
	\label{figS3}
\end{figure}

\begin{figure*}[t]
	\centering
	\includegraphics[width=\textwidth]{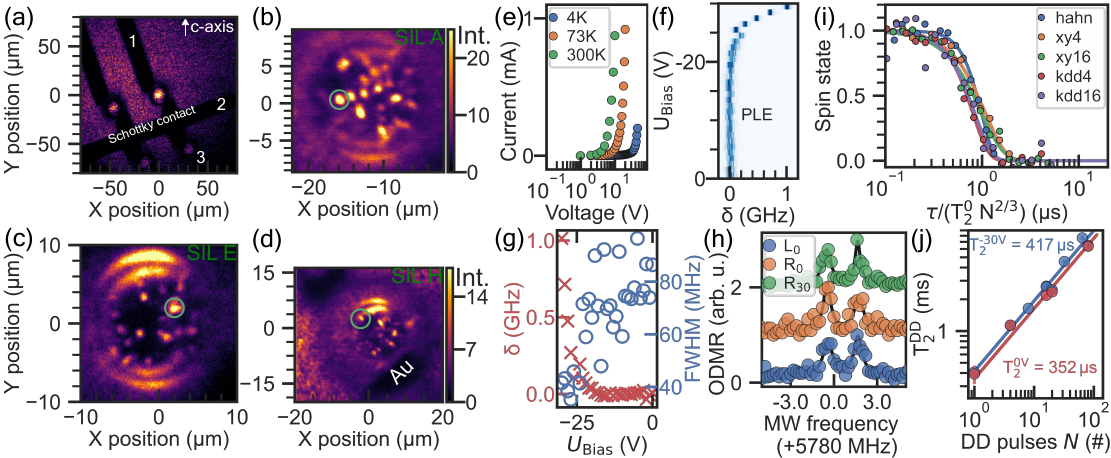}
	\caption{
		More detailed description of sample B:
		\textbf{(a)} Multiple SILs and the surrounding electrical contact.
		\textbf{(b)} SIL A, showing the V2 center which is coupled to the 5 MHz nuclear spin. Used for the DD sequence measurements. 
		\textbf{(c)} SIL E, showing the defect with three charge states.
		\textbf{(d)} SIL H, showing the defect with 2 charge states used for time transient plot.
		\textbf{(e)} I-V characteristics of contact 2 (in between the SILs)
		\textbf{(f)} Typical PLE scans under bias voltage.
		\textbf{(g)} Fitted linewidth and PLE position from (a).
		\textbf{(h)} ODMR measured at both transitions (L: left, R: right) from defect in (d) (main Fig 2e), showing that the PLE lines belong to the same defect. 
		ODMR measured also at -30 V bias, showing that there is no voltage dependency.
		\textbf{(i)} To $\tau$/T$_2$ rescaled DD sequences.
		\textbf{(j)} Comparsion of coherence times of the DD sequences for \SI{0}{V} and \SI{-30}{V}. Lines are fitted with $~N^{2/3}$ characteristics.
		}
	\label{figS4}
\end{figure*}

Large off-resonantly excited confocal scans have been taken at different bias voltages to show there is no difference in respect to each other, also by subtracting them no clear difference can be observed (Fig. \ref{figS3}). 
It seems the defect are enterly charge stable as no V2 appears or disappears. 
This is a contrary result to what has been observed previously in \cite{widmann2019electrical} at room temperature in a PIN diode.
\newline
Fig. \ref{figS4} shows additional information for sample 2.
There, confocal maps of the electrode geometry as well as of each SIL for every defect used in this paper. 
The voltage is usually applied between one of the top gold contacts (1-3) and the bottom of the sample connected to GND. 
A typical diode characteristics of contact 2 is shown in Fig. \ref{figS4}e for three temperatures.
For negative bias PLE Stark shift and linewidth narrowing down to 40 MHz can be observed (see Fig. \ref{figS4}f \& g), showing similar behaviour to previous measurements.
From the main text, the spin coherences at -30 V for different DD sequences are normalised to their number of refocussing pulses in Fig. \ref{figS4}f. 
Here, small deviations in from theoretical $\propto N^{2/3}$ dependency and fluctuations of the strechting exponent causing the non-identical lines.
For the sake of completeness, we also measured the spin coherence of one nuclear spin.
For the 5 MHz coupled $^{13}$C nuclear spin we observe a $T_2$ time of 63$\pm$4 ms.

\begin{figure*}[t]
	\centering
	\includegraphics[width=\textwidth]{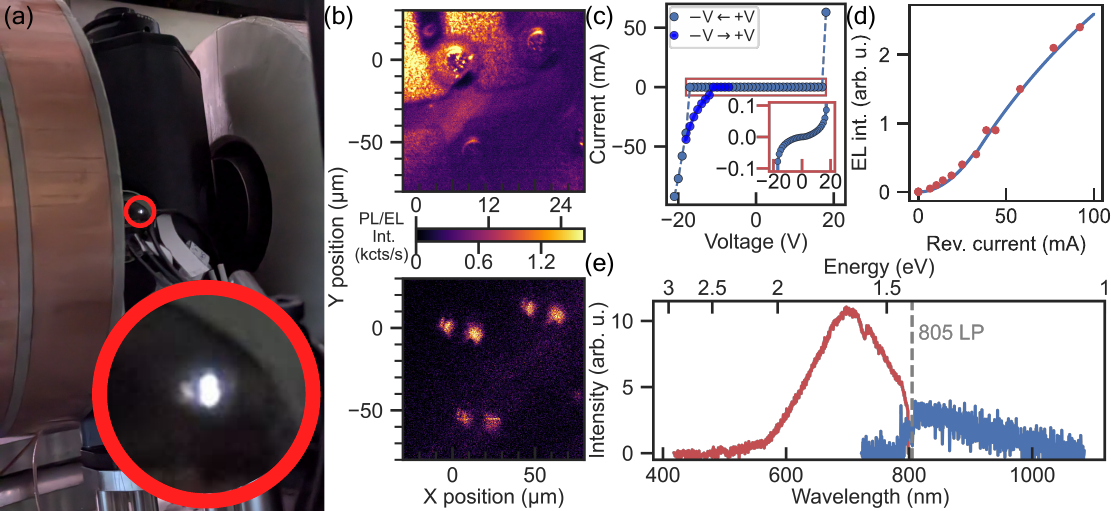}
	\caption{
		Electroluminescence in 4H-SiC at $\sim$\SI{10}{K}:
		\textbf{(a)} Picture of the Montana Cryostat, where the light emitting sample can be seen by pure eye. Luminescence appear white.
		\textbf{(b)} Confocal maps measured with PL and EL. The light emitted sidewise from the sample is reflected internally and is propagated to the objective after reflection at the SIL surfaces.
		\textbf{(c)} UI characteristics of contact 1 shows diode behaviour in both directions, positive and negative.
		For voltages beyond the threshold sudden jump in current can be observed. Sample heating up to \SI{14}{K} is caused by current flow. Dark blue curve shows hysteresis while going back to \SI{0}{V}. Red curve is a zoom in into the blue curve shown by the red rectangular. Note different units for each color, respectively.
		\textbf{(d)} EL intensity measured with the reflection signal on the upper left SIL show almost linear increase with rising current.
		\textbf{(e)} Spectra of the EL light, measured with two spectrometers on each side of the fixed \SI{805}{nm} LP filter. Shows really broad spectrum around \SI{700}{nm}.
		}
	\label{figS5}
\end{figure*}

As we can send current through a semiconductor host material if we pass the threshold voltage, we expect electro luminescence response of the material.
And indeed for forward voltages higher than 19 V we measure a diode current of several tens of milliampere. 
Besides of the rising base temperature of the cryo (increase by a few Kelvin), we observe a rising detected photon count rate on the photodetectors (SNSPD) with all laser sources being switched off.
The electroluminescence is strong enough that it is visible by eye as a small white light source inside of the cryotation (observable through a side window, see picture in Fig. \ref{figS5}a).
It seems that the light is almost emitting purely lateral, as it can be observed from the window but not from the objective on top. 
Only on the curved surface of the SILs some light can be detected (compare Fig. \ref{figS5}b upper and lower part), as there will be an appropriate angle to reflect the light upwards.
The electroluminescence intensity seem to grow linearly with the current (or exponentially with the voltage, Fig. \ref{figS5}d).
Interestingly, we also observed hysteresis on a rather large voltage range.
In contrast to the sharp rise of the current after passing the threshold, the breakthrough can be maintained almost to half the breakthrough voltage on the way back to 0 V (dark blue line in Fig. \ref{figS5}c), presumably due to increased local temperature.
With two Ocean Optics spectrometers for different wavelength ranges (as the emitted light unavoidable splits at a 800 LP filter) we observed the full spectra of the electroluminescence, shown in Fig. \ref{figS5}e.
It shows a very broad spectrum around 700 nm with a long tail into the near-infrared.

\section{Theoretical approach}
\begin{figure*}
	\centering
	\includegraphics[width=0.99\textwidth]{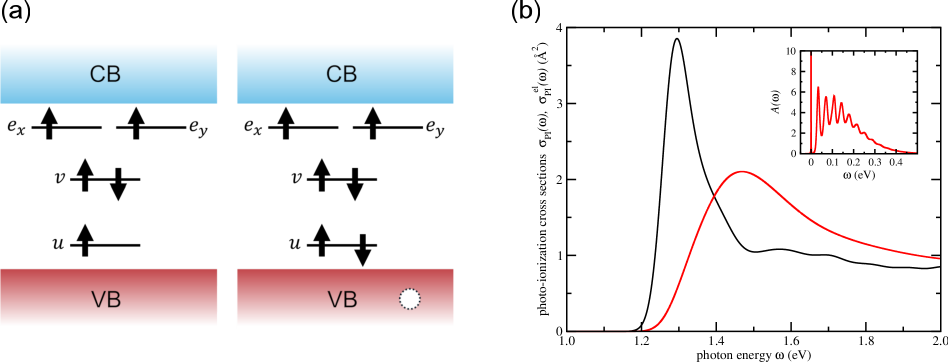}
	\caption{
		\textbf{(a)} Level schemes of the initial first excited quartet ($\text{V}_{\text{Si}}^{-}$) and	  final ionised states ($\text{V}_{\text{Si}}^{2-}+ h^{+}$).  
		In the PLE the excited quartet is produced from the groundstate by promoting an electron from the $u$ to $v$ level. 
		The second photon excites a valence band electron to the now empty $u$ level, generating the ground state of $\text{V}_{\text{Si}}^{2-}$ plus a	hole.
		\textbf{(b)} Photo-ionisation cross section ${\sigma}_{\text{PI}}^{\text{el}}$ (black) and $\sigma_{\text{PI}}$ (red). The spectral function between the first excited state and the ground state of $\text{V}_{\text{Si}}^{2-}$ used in the convolution (see Eq. \ref{eq:theory_SI_conv}) is shown in the inset.  
		}  
	\label{figS6}%
\end{figure*}

Here we outline the theoretical approach for the calculation of the photo-ionisation rate of the negatively charged cubic silicon vacancy $\text{V}_{\text{Si}}^{-}$ shown in Fig. 3d of the main text.  
In our experiments the defect is photo-ionised by a two photon process. 
The first photon resonantly excites the ground state to the first excited quartet state and subsequent photon with energy $h\nu$ from the laser with variable wavelength ionises the vacancy. 
In Fig. \ref{figS6}a we depict the process in which a doubly negative vacancy and a hole are produced, $\text{V}_{\text{Si}}^{-}\rightarrow\text{V}_{\text{Si}}^{2-}+h^{+}$.
We calculate the ionisation cross section from the resonantly excited state following \cite{razinkovas:21:2}.  
While the initial state $|\Psi_{\text{ex}}\rangle$ is resonantly excited from ground state and therefore does not contain any vibrational quanta, 
the final states $|\psi_{f,vib}\rangle$ describe the vibrationally excited ionised vacancy $\text{V}_{\text{Si}}^{2-}$ and a hole in the valence band. 
Within the Huang-Rhys theory, the cross section for the purely electronic transition between $|\Psi_{ex}\rangle$ and electronic part of the ionised states $|\Psi_{f}\rangle$ is given by
\begin{equation}
  \label{eq:sigma-el}
        \sigma_{\text{PI}}^{\text{el}}(\varepsilon)=\frac{4\pi^2\alpha}{3\,n_{D}}\varepsilon\sum_{ex,f} |\langle\Psi_{ex}|\hat{\vec{r}}|\Psi_f\rangle|^2\delta{(\varepsilon-\Delta E)}\ \ ,
\end{equation}
where $\Delta\/E=E_f-E_{ex}$, $\alpha$ is the fine-structure constant, and $n_{\text{D}}$ is the refractive index of 4H-SiC. 
The sum includes all spin substates of the first excited states and the final states with holes in the valence band that are compatible with the photon energy $\varepsilon=h\nu$.
Vibrational effects are incorporated into the full cross section $\sigma_{\text{PI}}(\varepsilon)$ by convolution of ${\sigma}_{\text{PI}}^{\text{el}}(\varepsilon)$ with the vibrational spectral function $A(\varepsilon)$ in the following way (cf. e.g. Ref. \cite{razinkovas:21:2})
\begin{equation}
	   \label{eq:theory_SI_conv}
    \sigma_{\text{PI}}(\varepsilon)=\varepsilon\,\int_{\infty}^{\infty}\,\frac{1}{\varepsilon'}\,
   \sigma_{\text{PI}}^{\text{el}}(\varepsilon')\,A(\varepsilon-\varepsilon')\,d\/\varepsilon'\ \ .
\end{equation}
The first step is the calculation of $\sigma_{\text{PI}}^{\text{el}}$. 
The states $|\Psi_{ex}\rangle$ and $|\Psi_{f}\rangle$ are correlated multiplet states and can be obtained by correlated first principles approaches such as our CI-CRPA method \cite{bockstedte:18,widmann2019electrical}. 
Only in special cases, these states reduce to single slater determinants and are directly accessible in (constrained) density functional theory \cite{bockstedte:18}. 
As the summation over the final states samples the k-space regarding the hole states and the convergence of this summation is demanding, the direct approach is computationally not feasible at present. 
Therefore we adopt a different approach and construct the initial and final multiplet states from ground state Kohn-Sham orbitals of $\text{V}_{\text{Si}}^{-}$ based on our CI-CRPA results for the cubic $\text{V}_{\text{Si}}^{-}$ and $\text{V}_{\text{Si}}^{2-}$. For instance, in the case of $M=\pm 3/2$ spin component of the initial first excited quartet (V2-line), 
the leading components ($\sim$\SI{94}{\%}) are given by $|u\/e_x\/e_y\rangle$ and $|\bar{u}\/\bar{e}_x\/\bar{e}_y\rangle$ using the hole notation and the overbar indicating negative spin.  
The second photon excites a hole $h$ ($\bar{h}$) into the empty $u$ ($\bar{u}$) yielding $|h\/e_x\/e_y\rangle$ ($|\bar{h}\/\bar{e}_x\/\bar{e}_y\rangle$), i.e. the triplet ground state of $\text{V}_{\text{Si}}^{2-}$ with a free hole $|^3A_2,M=-1\rangle\otimes|h\rangle=|e_x\/e_y\rangle\otimes|h\rangle$ ($|^3A_2,M=1\rangle\otimes|\bar{h}\rangle=|\bar{e}_x\/\bar{e}_y\rangle\otimes|\bar{h}\rangle$).
The transition matrix element in Eq. (\ref{eq:sigma-el}) becomes the matrix element between the Kohn-Sham states $|\langle h|\vec{r}|v\rangle|^2$, where we use the spin-orbitals from our spin-polarized calculation that are actually involved in theexcitation.

A similar analysis for the $M=\pm 1/2$ components of the first excited quartet, yields the same contributions to the cross section. 
The excited hole quartet state decomposes e.g. for $M=1/2$ into $\sqrt{\frac{2}{3}}\,|^3A_2,0\rangle\otimes|h\rangle$ and $\frac{1}{\sqrt{3}}\,|^3A_2,1\rangle\otimes|\bar{h}\rangle$, 
i.e. components of the ground state triplet of $\text{V}_\text{Si}^{2-}$ with a hole in the valence band. The excitation energy $\Delta\/E=E_f-E_{\text{ex}}$ is approximated by $\varepsilon(-|2-)-E_{\text{V2}}-\Delta E_h$, 
where $\varepsilon(-|2-)$, $E_{\text{V2}}$ and $\Delta E_h$ are the threshold for ionisation between the ground states of the two charge states with respect to the valence band edges, the transition energy of the $V2$ line, 
and the energy separation of the hole state from the valence band edge. 
For $\varepsilon(-|2-)$ we use the value from \cite{widmann2019electrical} obtained by hybrid DFT (see Fig.\ 3c). 
$E_{\text{V2}}$ is given by the experimental value which is well approximated by constrained DFT, and $\Delta E_h$ is taken from the DFT-PBE calculations (see below).

For the evaluation of the cross section $\sigma_{\text{PI}}^{\text{el}}$ (cf. Fig. \ref{figS6}b), we employ the projector augmented wave method as implemented in the VASP package \cite{kresse:96,kresse:99} and the PBE exchange-correlation functional \cite{perdew:96}. 
The defect is modeled using a hexagonal 6$\times$6$\times$2 supercell with 576 lattice sites. 
The Brillouin zone is sampled by a 15$\times$15$\times$15 k-point grid centered at the $\Gamma$ point following \cite{razinkovas:21:2}.  
For the geometry optimization of the ground state and the first excited quartet state (within constrained density functional theory), the calculation of ionisation levels, and to generate the starting point for calculations with the CI-CRPA method (cf.~\cite{bockstedte:18,
widmann2019electrical} for details) we employed the same supercells, the hybrid functional HSE \cite{heyd:03,krukau:06}, and a single k-point ($\Gamma$).  
For the description of the wave function a cut-off energy for of \SI{420}{eV} for the plane-waves and of \SI{840}{eV} for the augmentation is used. 
ionisation energies are calculated using charge state corrections (see \cite{widmann2019electrical}).
We obtain the spectral function $A(\varepsilon)$ (cf. the inset in Fig\ref{figS6}b) required for the full cross section $\sigma_{\text{PI}}(\varepsilon)$ via the generating function approach (e.g. \cite{razinkovas:21:1}). 
The reference configurations for the initial and final states are the equilibrium geometries of the first excited quartet $\text{V}_{\text{Si}}^-$ and the ground state of $\text{V}_{\text{Si}}^{2-}$ as in the final state the interaction of the free hole with the defect should be negligible.  
We perform the relaxation with the HSE exchange-correlation functional, in the former case within the constrained density functional theory and the occupation scheme depicted in Fig. \ref{figS6}a.
For the calculation of the vibrational spectrum, we employ the supercell embedding technique for the ground state of $\text{V}_{\text{Si}}^-$ and based on a 13$\times$13$\times$13 supercell and cut-offs $r_1=7.775$\,\AA~and $r_2=7.77$\,\AA~for the defect and bulk force constants, respectively. 
Force constants are calculated for a smaller 5$\times$5$\times$2 supercell, a 2$\times$2$\times$2 $\Gamma$-centered k-point set, and the PBE \cite{perdew:96} exchange-correlation functional. 
The bulk contributions are calculated with the PHONOPY package \cite{togo:23:1,togo:23:2}. 
To reduce the high computational costs associated with the defect cells, we employ the HIPHIVE toolkit \cite{eriksson:19,fransson:20} to fit the force constants to 65 rattled reference structures with the least-squares method. 
Our model achieves an RMSE of 9.2\,meV/\AA~when evaluated on a holdout test set containing \SI{5}{\%} of the dataset.

The theoretical power-dependent photon absorption rate $\Gamma_{\text{PI}}$  shown in (Fig.\ 3d) of the main text is obtained by multiplication of the cross section $\sigma_{\text{PI}}$ with the impinging photon flux $\Phi$ divided by the laser power $P$,
i.e.\ $\Gamma_{\text{PI}}(\lambda)=\sigma_{\text{PI}}(\frac{h\/c}{\lambda}) \,\frac{\lambda}{h\/c\,A_{\text{spot}}}$, where $A_{\text{spot}}=\pi\,(0.9 \lambda/N_{\text{A}})^2$ is the size of the laser spot for a numerical aperture of the focusing lens $N_{\text{A}}=0.75$.
Note that $\Gamma_{\text{PI}}$ in contrast to the measured two-photon ionisation rate does not include the quantum efficiency of the resonant PLE as well as competing relaxation processes that deplete the occupation of the first excited state. 
Such processes include spontaneous emission and spin-selective non-radiative relaxation via inter-system crossing and metastable states.  
In Fig. 3d, we therefore show $\Gamma_{\text{PI}}$ divided by its maximum value of $\Gamma_{\text{PI}}^{\text{max}}=$\SI{27873.13}{Hz/\micro W}.

%

